\documentclass{article}
\usepackage{amsmath,amsfonts,amssymb,amsthm,latexsym,stmaryrd}
\usepackage[]{caption2}
\usepackage[dvips]{graphicx,epsfig}
\usepackage{bbold}
\usepackage{setspace}
\usepackage[left=2.3cm,right=2.3cm,bottom=2.3cm,top=2.3cm]{geometry}

\newcommand{\ket}[1]{\mathop{\left|#1\right>}\nolimits}           

\begin{document}

\title{Nonlinear coherent states for the Susskind-Glogower operators}
\author{Roberto de Jes\'us Le\'on-Montiel, Hector Manuel Moya-Cessa, Francisco Soto-Eguibar\\
            \small Instituto Nacional de Astrof\'{\i}sica, \'{O}ptica y Electr\'{o}nica (INAOE)\\
            \small Luis Enrique Erro 1, Santa Mar\'{\i}a Tonantzintla, Puebla, 72840 Mexico\\
            \\
            \small Published as: REVISTA MEXICANA DE F\'{I}SICA S 57 (3) 133-147}

\maketitle

\begin{abstract}
We construct nonlinear coherent states for the Susskind-Glogower operators
by the application of the displacement operator on the vacuum state. We also
construct nonlinear coherent states as eigenfunctions of a Hamiltonian
constructed with the Susskind-Glogower operators. We generalize the solution
of the eigenfunction problem to an arbitrary $\mathop{\left|m\right>}
\nolimits$ initial condition. To analyze the obtained results, we plot the
Husimi $\mathcal{Q}$ function, the photon number probability distribution
and the Mandel $Q$-parameter. For both cases, we find that the constructed
states exhibit interesting nonclassical features, such as amplitude squeezing
and quantum interferences due to a self-splitting into two coherent-like
states. Additionally, we show that nonlinear coherent states may be modeled
by propagating light in semi-infinite arrays of optical fibers. \\

 \vspace{-4pt} \\

\end{abstract}

\section{Introduction}
Over the years, major effort has been directed towards the
generation of nonclassical states of  electromagnetic fields, in
which certain observables exhibit less fluctuations  than in a
coherent state, whose noise is referred to as the standard quantum
limit (SQL). Nonclassical states that have attracted the greatest
interest include (a) macroscopic quantum superpositions of
quasiclassical coherent states with different mean phases or
amplitudes, also called "Schr\"odinger cats" \cite{1,2}, (b)
squeezed states \cite{3}, whose fluctuations in one quadrature or the
amplitude are reduced beyond the SQL, and (c) the particularly
important limit of extreme squeezing, i.e. Fock or number states \cite{4}. \\

In this frame, the coherent states for the electromagnetic field, introduced by Glauber \cite{5,6} and Sudarshan \cite{7}, have been very important. Besides, their importance is increased by the fact that these states are relatively easy to produce in the laboratory and in their classical wave behavior. These states can be obtained by different mathematical definitions: \\
a) As the right-hand eigenstates of the annihilation operator \cite{6}
\begin{equation*}
\hat{a}\left\vert \alpha \right\rangle =\alpha \left\vert \alpha
\right\rangle ,
\end{equation*}%
with $\alpha $ a complex number. \newline
b) As the those states obtained by the application of the displacement
operator $\hat{D}\left( \alpha \right) =e^{\alpha \hat{a}^{\dagger }-\alpha
^{\ast }\hat{a}}$ on the vacuum state of the harmonic oscillator \cite{6}
\begin{equation*}
\left\vert \alpha \right\rangle =\hat{D}\left( \alpha \right) \left\vert
0\right\rangle .
\end{equation*}%
\newline
c) As those states which their time-evolving wave function shape does not
change with time and whose centroid follows the motion of a classical point
particle in a harmonic oscillator potential \cite{8}.

Using the harmonic oscillator algebra, one can obtain the same coherent
states from the three different definitions; however, for systems with
complex dynamical properties, the harmonic oscillator model is not adequate,
therefore new methods to generalize the idea of coherent states for systems
like those have been proposed.

Nieto and Simmons \cite{9,10,11} constructed coherent states
for potentials whose energy spectra have unequally spaced energy levels,
such as the Poschl-Teller potential, the harmonic oscillator with
centripetal barrier and the Morse potential.

Gazeau and Klauder \cite{12} proposed a generalization for systems with one
degree of freedom possessing discrete and continuous spectra. These states
were constructed performing a parametrization of the coherent states by two
real values: an amplitude $J$, and a phase $\gamma $, instead of a complex
value $\alpha $ .

Man'ko\textit{\ et al.}\cite{13} introduced coherent states of an $f$%
-deformed algebra as eigenstates of the annihilation operator $A=\hat{a}%
f\left( \hat{n}\right) $ where $\hat{n}=\hat{a}^{\dagger }\hat{a}$ is the
number operator and $\hat{a}$, $\hat{a}^{\dagger }$ are the annihilation and
creation boson operators of the harmonic oscillator algebra, respectively. A
remarkable result is that these states present nonclassical properties such
as squeezing and antibunching \cite{14}.

R\'ecamier\textit{\ et al.}\cite{15} proposed coherent states using a
deformed version of the displacement operator method generalized to the case
of $f$-deformed oscillators, assuming that the number operator function
appearing in the commutator between the deformed operators can be replaced
by a number. The method yields a displacement operator which is
approximately unitary and displaces the deformed annihilation $\hat{A}$ and
creation $\hat{A}^{\dagger }$ operators in the usual way.

In this work, we construct nonlinear coherent states by the application of
the displacement operator (for the Susskind-Glogower operators\cite{16}) on
the vacuum state and time-dependent $\left\vert m\right\rangle $ displaced
number states as eigenfunctions of a Hamiltonian representing the
fundamental physical coupling to the radiation field via the
Susskind-Glogower operators. We show the Husimi $\mathcal{Q}$ function \cite
{17} for the resulting states, as well as their photon distribution and
Mandel $Q$-parameter \cite{18}, in order to determine the nonclassical
features of the constructed states. Additionally, we show that nonlinear
coherent states may be modeled by propagating light in semi-infinite arrays
of optical fibers.

Nonlinear coherent states may be constructed using the Susskind-Glogower (%
\textbf{SG}) operators, for instance, defining a displacement operator for
them acting on the vacuum state. The Susskind and Glogower proposed
operators are
\begin{equation}  \label{sga}
\hat{V} =\sum_{n=0}^{\infty} \left\vert n \right>\left< n+1
\right \vert = \frac{1}{\sqrt{\hat{n}+1}}\hat{a} ,
\end{equation}
\begin{equation}  \label{sgc}
\hat{V}^{\dagger} = \sum _{n=0}^{\infty} \left\vert n+1
\right>\left< n \right \vert = \hat{a}^{\dagger}\frac{1}{\sqrt{\hat{n}+1}} ,
\end{equation}
satisfying the conditions
\begin{equation}
\hat{V} \left\vert n \right> = \left\vert n-1 \right> ,
\end{equation}
\begin{equation}
\hat{V}^{\dagger}\left\vert n \right> = \left\vert n+1 \right> .
\end{equation}
Additionally, we would like to make explicit the result
\begin{equation}
\hat{V} \left\vert 0 \right> = 0 ,
\end{equation}
that comes naturally from (\ref{sga}).

The \textbf{SG} operators possess a non-commuting and non-unitary nature,
that resides in the expressions
\begin{equation}
\hat{V}\hat{V}^{\dagger }=1,
\end{equation}%
\begin{equation}
\hat{V}^{\dagger }\hat{V}=1-\left\vert 0\right\rangle \left\langle
0\right\vert .
\end{equation}%
From the above expressions we can see that the non-commuting and non-unitary
nature of \textbf{SG} operators is only apparent for states of the radiation
field that have a significant overlap with the vacuum
\begin{equation}
\left\langle \psi \left\vert \left[ \hat{V},\hat{V}^{\dagger }\right]
\right\vert \psi \right\rangle =\left\langle \psi |0\right\rangle
\left\langle 0|\psi \right\rangle .
\end{equation}%
Therefore, for states where the vacuum contribution is negligible, we can
consider them as unitary and commutative and we can perform the following
approximation
\begin{equation}
\hat{V}^{-1}\simeq \hat{V}^{\dagger }.
\end{equation}%
The properties of the \textbf{SG} operators play an important role in the
development of the present work. For instance, if we analyze (\ref{sga}) and
(\ref{sgc}), we find that \textbf{SG} operators have the same form of the
ones we need to construct nonlinear coherent states, i.e., $\hat{V}=f\left(
\hat{n}+1\right) \hat{a}$ \hspace{1mm}and \hspace{1mm} $\hat{V}^{\dagger }=%
\hat{a}^{\dagger }f\left( \hat{n}+1\right) $ .

Following R\'ecamier \textit{et al.}\cite{15} we define \textbf{SG} coherent
states (from this point, we will refer to these states as \textbf{SG}
coherent states, keeping in mind that, in fact, they present a nonlinear
behavior) by
\begin{equation}
\left\vert \alpha \right\rangle _{\text{SG}}=e^{x\left( \hat{V}^{\dagger }-
\hat{V}\right) }\left\vert 0\right\rangle ,\text{ \ }x\in \mathbb{R}.
\label{5.1}
\end{equation}
As we have seen, commutation relations for the \textbf{SG} operators are not
simple, so we cannot factorize the displacement operator in a simple way,
which is why we propose two methods to solve the displacement operator. \\
First, we use the approximation $\hat{V}^{-1}\simeq \hat{V}^{\dagger }$ to
solve, approximately, the displacement operator. This solution helps us to
understand how the exact solution for the displacement operator should be.
Second, we solve the displacement operator in an exact way by developing it
in a Taylor series, that allows us to introduce the exact solution for
nonlinear coherent states, constructed with the \textbf{SG} operators.
Finally, we analyze the constructed states via the $\mathcal{Q}$ function
\cite{17}, the photon number distribution and the Mandel $Q$-parameter \cite
{18} in order to show their nonclassical features such as amplitude
squeezing and quantum interferences.

\section{Approximated displacement operator}
A first approach to factorize the displacement operator in the product of
exponentials, is to consider the approximation $\hat{V}^{-1}\simeq \hat{V}
^{\dagger }$. \\
Let us write the displacement operator as
\begin{equation}
D_{\text{SG}}\simeq e^{x\left( \hat{V}^{\dagger }-\left[ \hat{V}^{\dagger }
\right] ^{-1}\right) }.
\end{equation}
We find that the right-hand side of this equation corresponds to the
generating function of Bessel functions, that implies that
\begin{equation}
D_{\text{SG}}\simeq \sum_{n=-\infty }^{\infty }\hat{V}^{\dagger
n}J_{n}\left( 2x\right) ,
\end{equation}
where $J_{n}$ is the Bessel function of the first kind and order $n$. \\
Applying the displacement operator on the vacuum state, we have
\begin{equation}
D_{\text{SG}}\left\vert 0\right\rangle \simeq c_{0}\sum_{n=-\infty }^{\infty
}\hat{V}^{\dagger n}J_{n}\left( 2x\right) \left\vert 0\right\rangle ,
\end{equation}
and using that $\hat{V}^{\dagger }\left\vert n\right\rangle =\left\vert
n+1\right\rangle $ and that $\hat{V}^{-1}\simeq \hat{V}^{\dagger },$ we obtain
\begin{equation}
\left\vert \alpha \right\rangle _\text{SG}=D_\text{SG}\left\vert 0\right\rangle \simeq
c_{0}\sum_{n=0}^{\infty }J_{n}\left( 2x\right) \left\vert n\right\rangle .
\label{5.5}
\end{equation}
From the normalization requirement we determine $c_{0}$,
\begin{eqnarray}
  \nonumber_{\text{SG}}\left\langle \alpha |\alpha \right\rangle _{\text{SG}
}&=&c_{0}^{2}\sum_{n=0}^{\infty }J_{n}\left( 2x\right) \left\langle
n\right\vert \sum_{m=0}^{\infty }J_{m}\left( 2x\right) \left\vert
m\right\rangle \\
  &=&c_{0}^{2}\sum_{n=0}^{\infty }J_{n}^{2}\left( 2x\right) =1.
\end{eqnarray}
Using the result \cite{19}
\begin{equation}
1=J_{0}^{2}\left( 2x\right) +2\sum_{n=1}^{\infty }J_{n}^{2}\left( 2x\right) ,
\end{equation}
we obtain
\begin{equation}
c_{0}=\sqrt{\frac{2}{1+J_{0}^{2}\left( 2x\right) }},
\end{equation}%
and substituting in (\ref{5.5}), we finally get
\begin{equation}
\left\vert \alpha \right\rangle _\text{SG}\simeq \sqrt{\frac{2}{1+J_{0}^{2}\left(
2x\right) }}\sum_{n=0}^{\infty }J_{n}\left( 2x\right) \left\vert
n\right\rangle .  \label{5.9}
\end{equation}
As we mentioned before, solution (\ref{5.9}) helps us to foresee the exact
solution for these states. We can expect that the exact solution for \textbf{%
SG} coherent states corresponds to a linear combination of number states
where the coefficients are, except for some terms, Bessel functions of the
first kind and order $n$.

\section{Exact solution for the displacement operator}
In order to \textquotedblleft disentangle\textquotedblright\ the
displacement operator in an exact way, we can develop the exponential (\ref
{5.1}) in a Taylor series and then to evaluate the terms $\left( \hat{V}
^{\dagger }-\hat{V}\right) ^{k}$. For instance, for $k=7$ we have
\begin{equation}
\begin{split}
&\left( \hat{V}^{\dagger }-\hat{V}\right) ^{7} =\\
&=:\left( \hat{V}^{\dagger }-\hat{V}\right) ^{7}:+\binom{7}{2}\left( \left\vert 1\right\rangle
\left\langle 0\right\vert -\left\vert 0\right\rangle \left\langle1\right\vert \right)  \\
&-\binom{7}{1}\left( \left\vert 3\right\rangle \left\langle 0\right\vert
-\left\vert 2\right\rangle \left\langle 1\right\vert +\left\vert
1\right\rangle \left\langle 2\right\vert -\left\vert 0\right\rangle
\left\langle 3\right\vert \right) \\
&+\binom{7}{0}(\left\vert 5\right\rangle \left\langle 0\right\vert
-\left\vert 4\right\rangle \left\langle 1\right\vert +\left\vert
3\right\rangle \left\langle 2\right\vert -\left\vert 2\right\rangle
\left\langle 3\right\vert +\left\vert 1\right\rangle \left\langle
4\right\vert -\left\vert 0\right\rangle \left\langle 5\right\vert ),
\end{split}
\end{equation}
where $:\hspace{2mm}:$ means to arrange terms in such a way that the powers
of the operator $\hat{V}$ are always at the left of the powers of the
operator $\hat{V}^{\dagger }$.\newline
From the definition of the \text{SG} coherent states (\ref{5.1}), we can write
\begin{equation} \begin{split}
&\left\vert \alpha \right\rangle _\text{SG} = \\
&=e^{-x\hat{V}}e^{x\hat{V}^{\dagger}}\left\vert 0\right\rangle +\sum_{k=0}^{\infty }\frac{x^{k}}{k!}\sum_{n=0}^{
\left[ \frac{k}{2}-1\right] }\left( -1\right) ^{n}\binom{k}{n}\left\vert k-2n-2\right\rangle \\
&=e^{-x\hat{V}}e^{x\hat{V}^{\dagger}}\left\vert 0\right\rangle +\hat{V}^{2}\sum_{k=0}^{\infty }\frac{x^{k}}{k!}\sum_{n=0}^{\left[ \frac{k}{2}-1\right] }\left( -1\right) ^{n}\binom{k}{n}\hat{V}^{2n}\hat{V}^{\dagger^{k}}\left\vert 0\right\rangle ,
\end{split} \end{equation}
where the square brackets in the sum stand for the floor function, which maps a real number to the largest previous integer. \\
We can rewrite the above equation as
\begin{equation}  \begin{split}
&\left\vert \alpha \right\rangle _\text{SG}= \\
&=e^{-x\hat{V}}e^{x\hat{V}^{\dagger}}\left\vert 0\right\rangle +\hat{V}^{2}\sum_{k=0}^{\infty }\frac{x^{k}}{k!}
\sum_{n=0}^{\infty }\left( -1\right) ^{n}\binom{k}{n}\hat{V}^{2n}\hat{V}^{\dagger ^{k}}\left\vert 0\right\rangle ,
\end{split} \end{equation}
where we have taken the second sum to $\infty $ as we would add only zeros.
We now exchange the order of the sums
\begin{equation} \begin{split}
&\left\vert \alpha \right\rangle _\text{SG}= \\
&=e^{-x\hat{V}}e^{x\hat{V}^{\dagger}}\left\vert 0\right\rangle +\hat{V}^{2}\sum_{n=0}^{\infty}\sum_{k=n}^{\infty }\frac{\left( -1\right) ^{n}x^{k}}{\left( k-n\right)!n!}\hat{V}^{2n}\hat{V}^{\dagger ^{k}}\left\vert 0\right\rangle.
\end{split}  \end{equation}
By setting $m=k-n$ and using that
\begin{equation}
\hat{V}^{2n}\hat{V}^{\dagger ^{m+n}}\left\vert 0\right\rangle=\hat{V}^{n}\hat{V}^{\dagger ^{m}}\left\vert 0\right\rangle,
\end{equation}
we get
\begin{equation} \begin{split}
&\left\vert \alpha \right\rangle _\text{SG}= \\
&=e^{-x\hat{V}}e^{x\hat{V}^{\dagger }}\left\vert 0\right\rangle +\hat{V}
^{2}\sum_{n=0}^{\infty }\sum_{m=0}^{\infty }\left( -1\right) ^{n}\frac{
x^{m+n}}{m!n!}\hat{V}^{n}\hat{V}^{\dagger ^{m}}\left\vert 0\right\rangle
\notag \\
&=e^{-x\hat{V}}e^{x\hat{V}^{\dagger }}\left\vert 0\right\rangle +\hat{V}
^{2}\sum_{n=0}^{\infty }\frac{\left( -x\hat{V}\right) ^{n}}{n!}%
\sum_{m=0}^{\infty }\frac{\left( x\hat{V}^{\dagger }\right) ^{m}}{m!}
\left\vert 0\right\rangle  \notag \\
&=e^{-x\hat{V}}e^{x\hat{V}^{\dagger }}\left\vert 0\right\rangle +\hat{V}
^{2}e^{-x\hat{V}}e^{x\hat{V}^{\dagger }}\left\vert 0\right\rangle ,
\end{split}  \end{equation}
to finally obtain
\begin{equation}
\ket \alpha_\text{SG}=\left( 1+\hat{V}^{2}\right) e^{-x\hat{V}}e^{x\hat{V}^{\dagger }} \ket 0 .
\end{equation}
Applying the exponential terms on the vacuum state, we have
\begin{equation}
\begin{split}
&\left\vert \alpha \right\rangle _\text{SG}=\left( 1+\hat{V}^{2}\right)
\sum_{n=0}^{\infty }J_{n}\left( 2x\right) \left\vert n\right\rangle \\
&=\sum_{n=0}^{\infty }J_{n}\left( 2x\right) \left\vert n\right\rangle
+\sum_{n=2}^{\infty }J_{n}\left( 2x\right) \left\vert n-2\right\rangle ,
\end{split}
\end{equation}
making $m=n-2$ in the second sum and performing the index change $n=k-1$, we obtain
\begin{equation}
\left\vert \alpha \right\rangle _\text{SG}=\sum_{k=1}^{\infty }\left[
J_{k-1}\left( 2x\right) +J_{k+1}\left( 2x\right) \right] \left\vert
k-1\right\rangle .
\end{equation}
Using the recurrence relation of the Bessel functions \cite{19}
\begin{equation}
xJ_{n-1}\left( x\right) +xJ_{n+1}\left( x\right) =2nJ_{n}\left( x\right) ,
\end{equation}
and changing again the summation index, we finally write
\begin{equation}
\left\vert \alpha \right\rangle _\text{SG}=\frac{1}{x}\sum_{n=0}^{\infty }\left(
n+1\right) J_{n+1}\left( 2x\right) \left\vert n\right\rangle .  \label{5.21}
\end{equation}
Equation (\ref{5.21}) is an important result because it constitutes a new
expression for nonlinear coherent states. It remains to analyze the behavior
of the constructed states in order to determine the nonclassical features
that nonlinear coherent states may exhibit. \newline
Before we proceed with the analysis, and since we will need this result
later, we make clear that, as we can verify from (\ref{5.21}), for $x=0$, we have
\begin{equation}
\left\vert \alpha \left( x=0\right) \right\rangle _\text{SG}=\left\vert
0\right\rangle .
\end{equation}
\section{\textbf{SG} coherent states analysis}
There are different ways to find out if the state we are constructing
resembles one that we already know. Here, we will use three different
methods: the Husimi $\mathcal{Q}$ function \cite{17}, the photon number
distribution and the Mandel $Q$-parameter \cite{18}.
\subsection{The $\mathcal{Q}$ function}
\begin{figure}[h]  \begin{center}
                 \epsfig{file=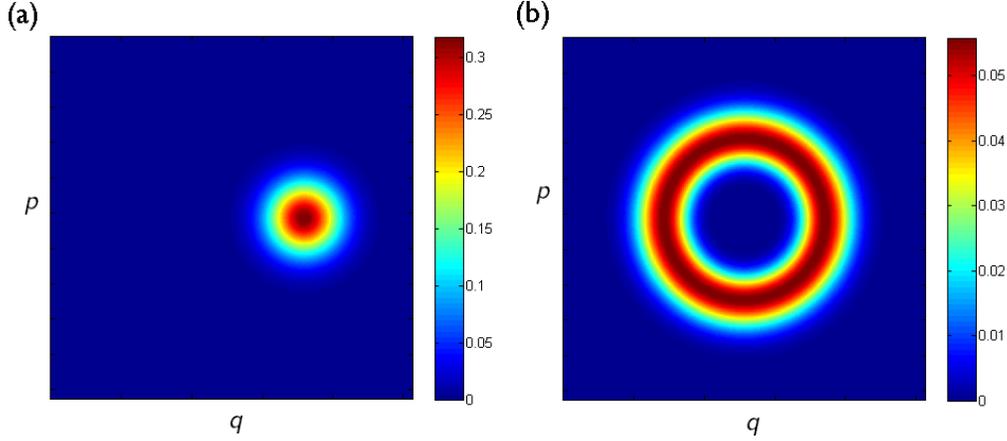,width=14cm}
                    \caption{{\protect\footnotesize {The $\mathcal{Q}$ function for
                   (a) coherent state $\ket \alpha $ with $\protect\alpha =2$ and (b) number state $\ket n $ with $n=5$.}}}
       \label{fig51}
\end{center} \end{figure}

\begin{figure}[h] \begin{center}
              \epsfig{file=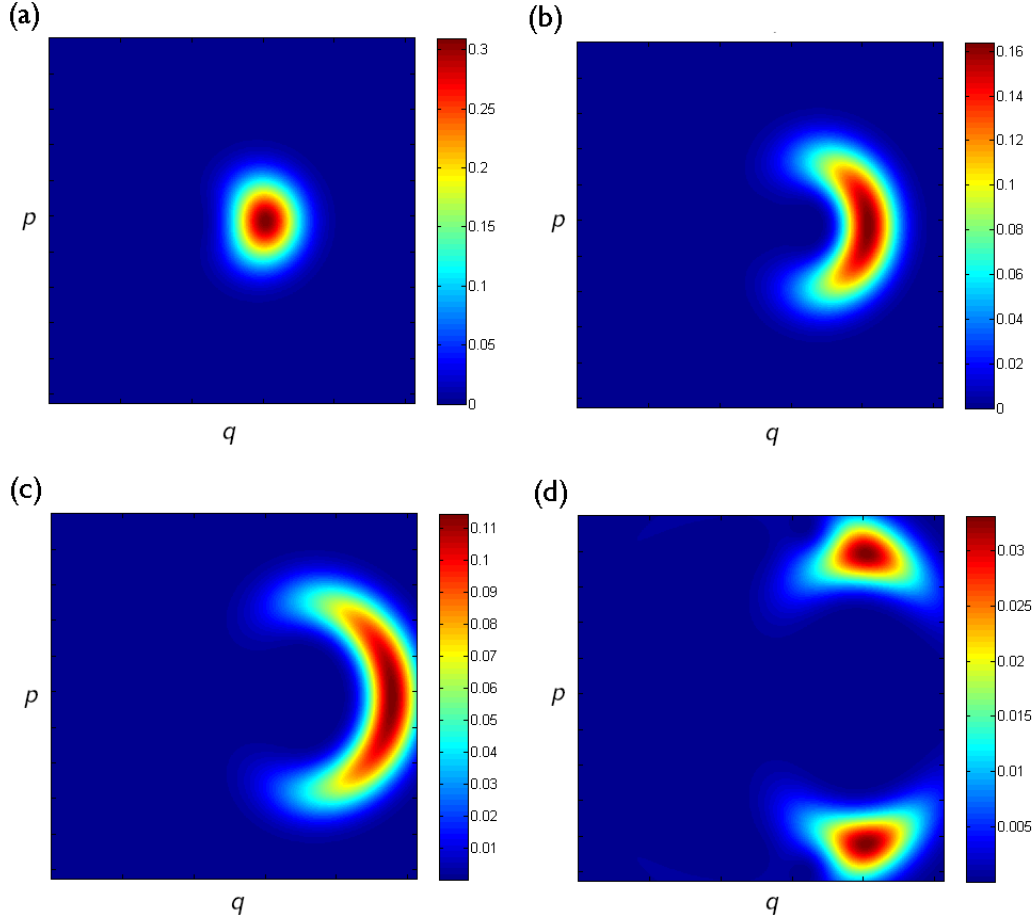,width=14cm}
                 \caption{{\protect\footnotesize {Exact \textbf{SG} coherent states $\mathcal{Q}$ function for \hspace{0.8mm}
                (a) $x=1$; \hspace{0.8mm}(b) $x=5$; \hspace{0.8mm}(c) $x=10$ \hspace{1mm} and \hspace{1mm}(d) $x=20$. }}}
      \label{fig52}
      \end{center} \end{figure}

The $\mathcal{Q}$ function, introduced by Husimi \cite{17}, corresponds to
a quasiprobability function that helps us to determine the behavior of a
quantum state in phase space. The $\mathcal{Q}$ function is defined as the
coherent state expectation value of the density operator and is given by
\begin{equation}
\mathcal{Q}=\frac{1}{\pi }\left\langle \alpha |\hat{\rho}|\alpha
\right\rangle .  \label{5.23}
\end{equation}
We plot the $\mathcal{Q}$ function for a coherent state $\ket \alpha$ and for a number state $\ket n$ in Figure \ref{fig51}.
If we substitute $\hat{\rho }=\mathop{\left|\psi \right>}\nolimits
\mathop{\left<\psi\,\right|}\nolimits$ for the \textbf{SG} coherent states
in (\ref{5.23}), we obtain
\begin{equation}
\mathcal{Q}_\text{SG}\left( \alpha ,t\right) =\frac{e^{-\left\vert \alpha
\right\vert ^{2}}}{\pi x^{2}}\left\vert \sum_{n=0}^{\infty }\frac{\alpha
^{\ast ^{n}}}{\sqrt{n!}}\left( n+1\right) J_{n+1}\left( 2x\right).
\right\vert ^{2}
\end{equation}
Figure \ref{fig52} shows the SG coherent states $\mathcal{Q}$ function for different
values of the parameter $x$.

We observe that the initial coherent state squeezes. Later, we will find the
value of $x$ for which we obtain the maximum squeezing of the coherent
state. We can also see that, as the parameter $x$ gets bigger, the state
splits into two coherent-like states, this produces quantum interferences,
as we will see later.

\subsection{Photon number distribution}
\begin{figure}[h]  \begin{center}
\epsfig{file=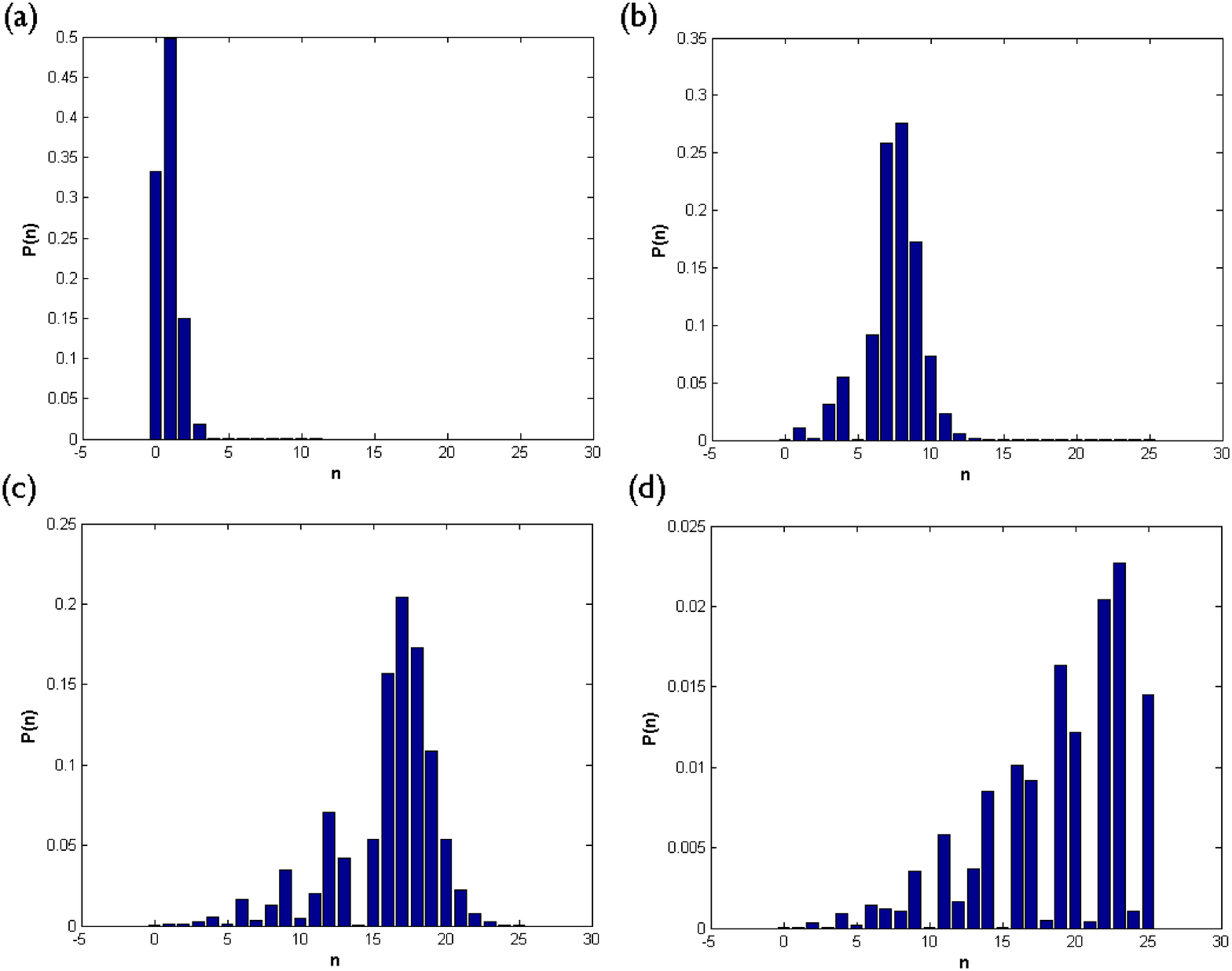,width=14cm}
\caption{{\protect\footnotesize {\textbf{SG} coherent states photon number
probability distributions for (a) $x=1$; (b) $x=5$; (c) $x=10$ and (d) $x=20$.}}}
      \label{fig54}
\end{center} \end{figure}

When one studies a quantum state, it is important to know about its photon
statistics. The photon number probability distribution $P\left( n\right)
=\langle n | \alpha \rangle \langle \alpha | n \rangle =\left\vert \langle n
| \alpha \rangle \right\vert ^{2}$ is useful to determine amplitude
squeezing. We should refer to amplitude squeezed light as light for which
the photon number distribution is usually narrower than the one of a
coherent state of the same amplitude. The photon number distribution is also
useful to analyze if there exist effects due to quantum interferences. Using
$P\left( n\right) =\left\vert \langle n | \alpha \rangle \right\vert ^{2}$,
we write the \textbf{SG} coherent states photon number distribution as
\begin{equation}
P\left( n\right) =\left\vert \frac{1}{x}\left( n+1\right) J_{n+1}\left(
2x\right) \right\vert ^{2}.
\end{equation}%
Figure \ref{fig54} shows the \textbf{SG} coherent states photon number distribution
for different values of the amplitude parameter $x$.

Figure \ref{fig54} helps us to understand the effect of quantum interferences; for
instance, consider Figure \ref{fig54}(c). We see that it is not a uniform distribution
of photons, the distribution has \textquotedblleft holes\textquotedblright ;
these holes are the consequence of the interference between the two states
arising from the splitting of the initial one. \newline
Comparing Figure \ref{fig54}(a) with the one obtained for a coherent state wave function
that is moving through the harmonic oscillator
potential between the classical turning points, we can see that the photon
number distribution for the \textbf{SG} coherent states is narrower than the
one for a coherent state of the same amplitude. This suggests that we are,
indeed, obtaining an amplitude squeezed state. It is interesting to know
when the state is maximally squeezed, but we need another tool to obtain the
value of the parameter $x$ for which this occurs.

\subsection{Mandel $Q$-parameter}

\begin{figure}[h]  \begin{center}
      \epsfig{file=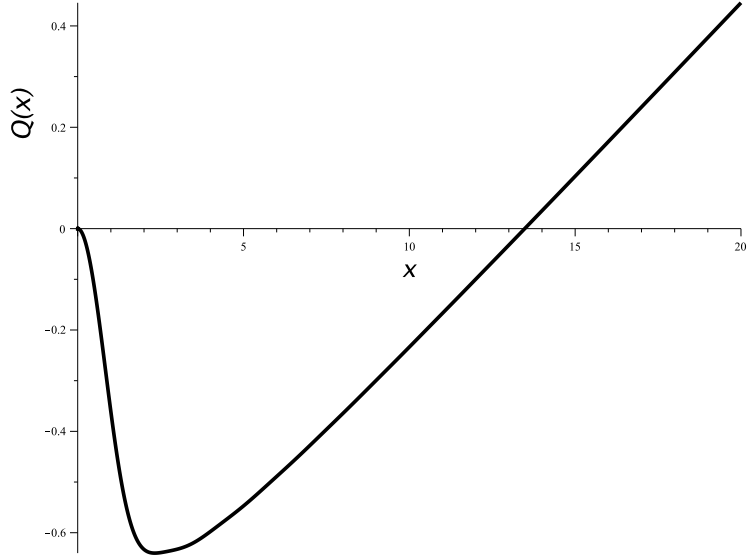,width=10cm}
      \caption{{\protect\footnotesize {Mandel $Q$-parameter for the \textbf{SG} coherent states.}}}
      \label{fig55}
\end{center}  \end{figure}

There has been an extensive argument about which is the better way to
determine the quantumness of a given state; nevertheless, there exists a
very useful tool for determining the nature of the states we have
constructed. This tool is called the Mandel $Q$-parameter \cite{18}. We will
use it not only because it represents a good parameter to define the
quantumness of \textbf{SG} coherent states but also, because it will allow
us to find the domain of $x$ for which they exhibit a nonclassical behavior;
moreover, it will help us to find the value of $x$ for which the state is
maximally squeezed. \newline
The Mandel $Q$-parameter is defined by
\begin{equation}
Q=\frac{\left\langle \hat{n}^{2}\right\rangle -\left\langle \hat{n}%
\right\rangle ^{2}}{\left\langle \hat{n}\right\rangle }-1,  \label{5.26}
\end{equation}%
where
\begin{equation}
\text{If}\hspace{2mm}Q\hspace{2mm}\left\{
\begin{array}{ll}
>0, & \text{super Poissonian distribution} \\
=0, & \text{Poissonian distribution (coherent state)} \\
<0, & \text{sub-Poissonian} \\
=-1, & \text{number state}.
\end{array}
\right.
\end{equation}
For the \textbf{SG} coherent states we have that
\begin{equation}
\left\langle \hat{n}\right\rangle =\frac{1}{x^{2}}\left[ \sum_{k=0}^{\infty
}k^{3}J_{k}^{2}\left( 2x\right) -\sum_{k=0}^{\infty }k^{2}J_{k}^{2}\left(
2x\right) \right] ,  \label{5.28}
\end{equation}
and
\begin{equation}
\begin{split}
&\left\langle \hat{n}^{2}\right\rangle = \\
&=\frac{1}{x^{2}}\left[ \sum_{k=0}^{\infty }k^{4}J_{k}^{2}\left( 2x\right)
-2\sum_{k=0}^{\infty }k^{3}J_{k}^{2}\left( 2x\right) +\sum_{k=0}^{\infty
}k^{2}J_{k}^{2}\left( 2x\right) \right] .  \label{5.29}
\end{split}%
\end{equation}%
\newline
The even sums, with respect to the power of $k$, in (\ref{5.28}) and (\ref{5.29}), can be evaluated. \\
In the Appendix, we show explicitly that
\begin{equation}
\sum_{k=0}^{\infty }k^{2}J_{k}^{2}\left( 2x\right) =x^{2},
\end{equation}
and
\begin{equation}
\sum_{k=0}^{\infty }k^{4}J_{k}^{2}\left( 2x\right) =3x^{4}+x^{2}.
\end{equation}
The sum $\sum_{k=0}^{\infty }k^{3}J_{k}^{2}\left( 2x\right) $ is more
complicated, and can be evaluated using the technics developed by Dattolli
\textit{et al} \cite{20}, to
\begin{equation}
\begin{split}
&\sum\limits_{k=1}^{\infty }k^{3}J_{k}^{2}\left( 2x\right)= \allowbreak
x^2 \{(6x^2+1)J_{0}^{2}(2x)+(6x^2-1)J_{1}^{2}(2x) \\
&-2xJ_{0}(2x)J_{1}(2x)+\frac{2x^2}{3}[J_{0}(2x)J_{2}(2x)+J_{1}(2x)J_{3}(2x)] \}.
\end{split}
\end{equation}
Substituting the values of the sums into equation (\ref{5.26}), we obtain the plot shown in Figure \ref{fig55}.
From Figure \ref{fig55} we can see that, depending on the parameter $x$, the photon
distribution of the constructed states is sub-poissonian, $Q<0$, meaning
that amplitude squeezing states may be find for a value of $x$ within the
domain $0<x\leq 13.48$. Also, we find that the most squeezed state may be
obtained at $x=2.32$, with $Q=-0.64$. \newline
At this point, all the comments on the results have been made considering $x$
as a parameter; however, in what follows, we will see that $x$ corresponds
to an expression that depends explicitly on what we should refer to as
interaction time.

\section{Eigenfunctions of the \textbf{SG} Hamiltonian}
Previously we managed to construct nonlinear coherent states applying the
displacement operator on the vacuum state. However, even when the obtained
results are very interesting, we cannot avoid to wonder how these states
could be physically interpreted. Physical interpretations are given by
operators representing observables, i.e., quantities that can be measured in
the laboratory. The most important observable is the Hamiltonian, this
operator helps us to find the energy distribution of an state via its
eigenvalues. \\

Here, we construct \textbf{SG} coherent states as eigenfunctions of a
Hamiltonian that we propose and which represents the fundamental coupling to
the radiation field via the \textbf{SG} operators. In section 5.1 we
construct time-dependent \textbf{SG} coherent states with the vacuum state $
\left\vert 0\right\rangle $ as initial condition, i.e., we construct states
that satisfy $\left\vert \alpha \left( x=0\right) \right\rangle _\text{SG}
=\left\vert 0\right\rangle $. In section 5.2 we generalize the eigenfunction
problem for an arbitrary $\left\vert m\right\rangle $ initial condition, we
also show that previous results correspond to the particular case $m=0$ and
in section 5.3 we make use of the three methods presented in the previous
chapter in order to analyze the properties and time-evolution of the
constructed states.

\subsection{Solution for $\left\vert 0 \right>$ as initial condition}
As we mentioned before, it is possible to construct \textbf{SG} coherent
states as eigenfunctions of the interaction Hamiltonian
\begin{equation}
\hat{H}=\eta \left( \hat{V}+\hat{V}^{\dagger }\right) ,  \label{6.1}
\end{equation}
where $\eta $ is the coupling coefficient. Hamiltonians like this may be produced in ion-traps \cite{21,22,23}. \\
The Hamiltonian proposed in (\ref{6.1}) corresponds to a variation of the
one used in \cite{23} to model physical couplings to the radiation field.
Here, physical couplings take place via the \textbf{SG} operators.\newline
We write the eigenfunctions of the Hamiltonian (\ref{6.1}) in the
interaction picture as
\begin{equation}
\mathop{\left|\psi \right>}\nolimits =\sum_{n=0}^{\infty }C_{n}%
\mathop{\left|n \right>}\nolimits
\end{equation}
and we have
\begin{equation}
\hat{H}\left\vert \psi \right\rangle =\eta \sum_{n=1}^{\infty
}C_{n}\left\vert n-1\right\rangle +\eta \sum_{n=0}^{\infty }C_{n}\left\vert
n+1\right\rangle .
\end{equation}
Now, changing the summation indexes, we obtain
\begin{equation}
\begin{split}
\hat{H}\left\vert \psi \right\rangle =\eta \sum_{n=0}^{\infty
}C_{n+1}\left\vert n\right\rangle +\eta \sum_{n=1}^{\infty
}C_{n-1}\left\vert n\right\rangle \\
=\eta C_{1}\left\vert 0\right\rangle +\eta \sum_{n=1}^{\infty }\left(
C_{n+1}+C_{n-1}\right) \left\vert n\right\rangle .
\end{split}
\end{equation}
Then
\begin{equation}
\begin{split}
&\eta C_{1}\left\vert 0\right\rangle +\sum_{n=1}^{\infty }\eta \left(
C_{n+1}+C_{n-1}\right) \left\vert n\right\rangle = \\
&=EC_{0}\left\vert 0\right\rangle +\sum_{n=1}^{\infty }EC_{n}\left\vert
n\right\rangle.
\end{split}
\end{equation}
Comparing coefficients with same number states of the sum, we have
\begin{equation}
C_{1}=\frac{E}{\eta }C_{0},
\end{equation}
\begin{equation}
\eta \left( C_{n+1}+C_{n-1}\right) =EC_{n}.
\end{equation}
These are the recurrence relation of the Chebyshev polynomials of the second
kind (\cite{19}), and we can write
\begin{equation}
\left\vert \psi \left( t;\xi \right) \right\rangle =\sum_{n=0}^{\infty
}e^{-iEt}U_{n}\left( \xi \right) \left\vert n\right\rangle ,  \label{6.10}
\end{equation}
where
\begin{equation}
E=2\eta \xi .
\end{equation}
However (\ref{6.10}) does not satisfies the initial condition $\left\vert
\psi \left( t=0\right) \right\rangle =\left\vert 0\right\rangle $. Moreover,
the solution (\ref{6.10}) has the parameter $\xi $ and, as we see from (\ref
{5.1}), we should not have another parameter, except for the time $t$. \\

A way to construct a solution, as the one we previously obtained in (\ref
{5.21}), is by looking at the exponential term in (\ref{6.10}) and noticing
that it has the form of the Fourier transform kernel; so, we need to propose
a $\xi $-dependent function and integrate it over all $\xi $, in order to
obtain a solution where time is the only variable. We have then,
\begin{equation}
\left\vert \psi \left( t\right) \right\rangle =\sum_{n=0}^{\infty
}\int_{-\infty }^{\infty }d\xi P\left( \xi \right) U_{n}\left( \xi \right)
e^{-i2\eta \xi t}\left\vert n\right\rangle .  \label{6.12}
\end{equation}%
We see from the above equation that $\left\vert \psi \left( t\right)
\right\rangle $ corresponds to a sum of Fourier transforms of Chebyshev
polynomials with respect to a weight function $P\left( \xi \right) $. This
kind of Fourier transforms may be solved by using the following result \cite{24}
\begin{equation}
\begin{split}
&\mathcal{F}\left\{ \frac{J_{n}\left( \omega \right) }{\omega }\right\} = \\
&=\sqrt{\frac{2}{\pi }}\frac{i}{n}\left( -i\right) ^{n}U_{n-1}\left( \xi
\right) \sqrt{1-\xi ^{2}}\text{rect}\left( \frac{\xi }{2}\right) ,
\label{6.13}
\end{split}%
\end{equation}
where $\mathcal{F}\left\{ {}\right\} $ is the Fourier transform and
\begin{equation}
\text{rect}\left( \frac{\xi }{2}\right) =\left\{
\begin{array}{ll}
1 & ,\hspace{1mm}-1\leq \xi \leq 1 \\
0 & \text{, otherwise.}%
\end{array}
\right.  \label{6.14}
\end{equation}
From (\ref{6.13}) we write
\begin{equation}
\begin{split}
          &\frac{J_{n}\left( \omega \right) }{\omega }= \\
          &=\frac{1}{\sqrt{2\pi }}\int_{-\infty }^{\infty } \\
          &\times \sqrt{\frac{2}{\pi }}\frac{i}{n}\left( -i\right) ^{n}U_{n-1}\left( \xi \right) \sqrt{1-\xi ^{2}}\text{
           rect}\left( \frac{\xi }{2}\right) e^{i\omega \xi }d\xi .  \label{6.16}
\end{split}
\end{equation}
Using definition (\ref{6.14}), and making $k=n-1$, we obtain
\begin{equation}
\frac{J_{k+1}\left( \omega \right) }{\omega }=\int_{-1}^{1}\left[ \frac{1}{%
\pi }\frac{i\left( -i\right) ^{k+1}}{k+1}\sqrt{1-\xi ^{2}}\right]
U_{k}\left( \xi \right) e^{i\omega \xi }d\xi .  \label{6.17}
\end{equation}%
With equation (\ref{6.17}) it is possible to solve the integral in (\ref%
{6.12}). Writing the weight function as
\begin{equation}
P\left( \xi \right) =\frac{2}{\pi }\sqrt{1-\xi ^{2}}\text{rect}\left( \frac{%
\xi }{2}\right) ,
\end{equation}%
substituting it into equation (\ref{6.12}) and using now (\ref{6.17}), we
get
\begin{equation}
\left\vert \psi \left( t\right) \right\rangle =\frac{2}{-2\eta t}%
\sum_{n=0}^{\infty }i^{n}\left( n+1\right) J_{n+1}\left( -2\eta t\right)
\left\vert n\right\rangle .
\end{equation}%
Considering the odd parity of the Bessel functions, we finally obtain
\begin{equation}
\left\vert \psi \left( t\right) \right\rangle =\frac{1}{\eta t}%
\sum_{n=0}^{\infty }i^{n}\left( n+1\right) J_{n+1}\left( 2\eta t\right)
\left\vert n\right\rangle .  \label{6.20}
\end{equation}%
We see that $\left\vert \psi \left( t\right) \right\rangle $ in equation (%
\ref{6.20}) depends only on $t$ and considering that
\begin{equation}
\lim_{2\eta t\rightarrow 0}\frac{J_{n}\left( 2\eta t\right) }{2\eta t}%
=\left\{
\begin{array}{ll}
0 & ,\hspace{1mm}n=2,3,4,... \\
&  \\
\frac{1}{2} & ,\hspace{3mm}n=1,%
\end{array}%
\right.
\end{equation}%
we can verify that
\begin{equation}
\left\vert \psi \left( t=0\right) \right\rangle =\left\vert 0\right\rangle .
\end{equation}%
The equation (\ref{6.20}) corresponds to the expression for \textbf{SG}
coherent states that we obtained previously (\ref{5.21}). \\

The solution presented in this section allows us to notice that, while in
the previous chapter $x$ was only a parameter, now it represents something
physical. It may be related to an interaction time, for example, in the
motion of a trapped atom \cite{23}. We have managed to construct the same
expression for the \textbf{SG} coherent states as the one obtained by the
application of the displacement operator on the vacuum state; however, we
will see that the formalism presented in this section may be used to
generalize the solution for an arbitrary initial condition $\left\vert
m\right\rangle $.

\subsection{Solution for $\left\vert m \right>$ as initial condition}
At this point, we have managed to construct \textbf{SG} coherent states,
first as those obtained by the application of the displacement operator on
the vacuum state and later, as eigenfunctions of the Hamiltonian (\ref{6.1}%
); however, they are a particular case of a more general expression. \newline
Using the recurrence relation for the Chebyshev polynomials and the result (%
\ref{6.17}), it is possible to generalize \textbf{SG} coherent states to an
arbitrary $\left\vert m\right\rangle $ initial condition, where $m=0,1,2,...$%
. \newline
From equation (\ref{6.12}), we have that
\begin{equation}
\left\vert \psi \left( t=0\right) \right\rangle =\sum_{n=0}^{\infty
}\int_{-\infty }^{\infty }d\xi P\left( \xi \right) U_{n}\left( \xi \right)
\left\vert n\right\rangle ,
\end{equation}
and considering the function
\begin{equation}
P_{m}\left( \xi \right) =\frac{2}{\pi }\sqrt{1-\xi ^{2}}U_{m}\left( \xi
\right) \text{rect}\left( \frac{\xi }{2}\right),
\end{equation}
and the well known Chebyshev polynomials of the second kind orthonormal condition
\begin{equation*}
\frac{2}{\pi }\int_{-1}^{1}U_{n}\left( \xi \right) U_{m}\left( \xi \right)
\sqrt{1-\xi ^{2}}d\xi =\delta _{nm},
\end{equation*}
we obtain
\begin{equation}
\left\vert \psi \left( t=0\right) \right\rangle =\sum_{n=0}^{\infty }\delta
_{nm}\left\vert n\right\rangle .
\end{equation}
To obtain the solution for the $m$-state, let us consider the particular
cases $m=0$ and $m=1$.\newline
For $m=0$,
\begin{equation}
\begin{split}
& \left\vert \psi \left( t\right) \right\rangle _{m=0}= \\
& =\sum_{n=0}^{\infty }\int_{-\infty }^{\infty }P_{0}\left( \xi \right)
U_{n}\left( \xi \right) e^{-i2\eta \xi t}d\xi \left\vert n\right\rangle  \\
& =\sum_{n=0}^{\infty }\int_{-\infty }^{\infty }\frac{2}{\pi }\sqrt{1-\xi
^{2}}U_{0}\left( \xi \right) U_{n}\left( \xi \right) e^{-i2\eta \xi t}\text{
rect}\left( \frac{\xi }{2}\right) d\xi \left\vert n\right\rangle  \\
& =\sum_{n=0}^{\infty }\int_{-1}^{1}\frac{2}{\pi }\sqrt{1-\xi ^{2}}%
U_{0}\left( \xi \right) U_{n}\left( \xi \right) e^{-i2\eta \xi t}d\xi
\left\vert n\right\rangle  \\
& =\sum_{n=0}^{\infty }\int_{-1}^{1}\frac{2}{\pi }\sqrt{1-\xi ^{2}}%
U_{n}\left( \xi \right) e^{-i2\eta \xi t}d\xi \left\vert n\right\rangle .
\end{split}%
\end{equation}%
Using (\ref{6.16}) and making $n\rightarrow n+1$, we have
\begin{equation}
\left\vert \psi \left( t\right) \right\rangle _{m=0}=\sum_{n=0}^{\infty
}i^{n}\left( n+1\right) \frac{J_{n+1}\left( 2\eta t\right) }{\eta t}%
\left\vert n\right\rangle .
\end{equation}%
Using the recurrence relation
\begin{equation}
xJ_{n-1}\left( 2x\right) +xJ_{n+1}\left( 2x\right) =nJ_{n}\left( 2x\right) ,
\label{6.28}
\end{equation}%
and changing the summation index, we finally obtain
\begin{equation}
\left\vert \psi \left( t\right) \right\rangle _{m=0}=\sum_{n=0}^{\infty }
\left[ i^{n}J_{n}\left( 2\eta t\right) +i^{n}J_{n+2}\left( 2\eta t\right) %
\right] \left\vert n\right\rangle .  \label{6.29}
\end{equation}
For $m=1$, we have
\begin{equation}
\begin{split}
& \left\vert \psi \left( t\right) \right\rangle _{m=1}= \\
& =\sum_{n=0}^{\infty }\int_{-\infty }^{\infty }P_{1}\left( \xi \right)
U_{n}\left( \xi \right) e^{-i2\eta \xi t}d\xi \left\vert n\right\rangle  \\
& =\sum_{n=0}^{\infty }\int_{-\infty }^{\infty }\frac{2}{\pi }\sqrt{1-\xi
^{2}}U_{1}\left( \xi \right) U_{n}\left( \xi \right) e^{-i2\eta \xi t}\text{
rect}\left( \frac{\xi }{2}\right) d\xi \left\vert n\right\rangle \\
& =\sum_{n=0}^{\infty }\int_{-1}^{1}\frac{2}{\pi }\sqrt{1-\xi ^{2}}
U_{1}\left( \xi \right) U_{n}\left( \xi \right) e^{-i2\eta \xi t}d\xi
\left\vert n\right\rangle \\
& =\sum_{n=0}^{\infty }\int_{-1}^{1}\frac{2}{\pi }\sqrt{1-\xi ^{2}}\left(
2\xi \right) U_{n}\left( \xi \right) e^{-i2\eta \xi t}d\xi \left\vert
n\right\rangle .
\end{split}
\end{equation}
Using the recurrence relation for the Chebyshev polynomials of the second
kind, we write
\begin{equation}
\begin{split}
& \left\vert \psi \left( t\right) \right\rangle _{m=1}= \\
& \sum_{n=0}^{\infty }\int_{-1}^{1}\frac{2}{\pi }\sqrt{1-\xi ^{2}}\left(
U_{n-1}\left( \xi \right) +U_{n+1}\left( \xi \right) \right) e^{-i2\eta \xi
t}d\xi \left\vert n\right\rangle .
\end{split}
\end{equation}
Considering (\ref{6.16}) for the first integral, and making $n=n+2$ for the second one, we have
\begin{equation}
\begin{split}
& \left\vert \psi \left( t\right) \right\rangle _{m=1}= \\
& =\sum_{n=0}^{\infty }\frac{1}{\eta t}\left[ i^{n+1}\left( n+2\right)
J_{n+2}\left( 2\eta t\right) +i^{n-1}nJ_{n}\left( 2\eta t\right) \right]
\left\vert n\right\rangle.  \label{6.31}
\end{split}
\end{equation}
We now construct a different expression for the above equation (\ref{6.31}).
Let us write it in the following way,
\begin{equation}
\begin{split}
& \left\vert \psi \left( t\right) \right\rangle _{m=1}= \\
& \sum_{n=0}^{\infty }\left\{ -i^{n+2-1}\frac{1}{\eta t}nJ_{n}\left( 2\eta
t\right) +i^{n+2-1}\frac{1}{\eta t}\left( n+2\right) J_{n+2}\left( 2\eta
t\right) \right\} ,
\end{split}
\end{equation}
making $k=n+2$
\begin{equation}
\begin{split}
& \left\vert \psi \left( t\right) \right\rangle _{m=1}= \\
& \sum_{n=0}^{\infty }\left\{ -i^{k-1}\frac{1}{\eta t}\left( k-2\right)
J_{k-2}\left( 2\eta t\right) +i^{k-1}\left[ \frac{1}{\eta t}kJ_{k}\left(
2\eta t\right) \right] \right\} ,
\end{split}
\end{equation}
using the recurrence relation (\ref{6.28}), we write
\begin{equation}
\begin{split}
& \left\vert \psi \left( t\right) \right\rangle _{m=1}=\allowbreak
\sum_{n=0}^{\infty }\left\{
\begin{array}{c}
-i^{k-1}\frac{1}{\eta t}\left( k-2\right) J_{k-2}\left( 2\eta t\right)  \\
+i^{k-1}\left[ J_{k-1}\left( 2\eta t\right) +J_{k+1}\left( 2\eta t\right)
\right]
\end{array}
\right\}  \\
& =\allowbreak \sum_{n=0}^{\infty }\left\{
\begin{array}{c}
-i^{k-1}\left[ \frac{1}{\eta t}\left( k-2\right) J_{k-2}\left( 2\eta
t\right) -J_{k-1}\left( 2\eta t\right) \right]  \\
+i^{k-1}J_{k+1}\left( 2\eta t\right)
\end{array}
\right\}  \\
& =\allowbreak \sum_{n=0}^{\infty }\left\{
\begin{array}{c}
i^{k-3}\left[ \frac{1}{\eta t}\left( k-2\right) J_{k-2}\left( 2\eta t\right)
-J_{k-1}\left( 2\eta t\right) \right]  \\
+i^{k-1}J_{k+1}\left( 2\eta t\right)
\end{array}
\right\}  \\
& =\allowbreak \sum_{n=0}^{\infty }\left\{
\begin{array}{c}
i^{k-2-1}\left[ \frac{1}{\eta t}\left( k-2\right) J_{k-2}\left( 2\eta
t\right) -J_{k-2+1}\left( 2\eta t\right) \right]  \\
+i^{k-2+1}J_{k-2+3}(2\eta t)
\end{array}
\right\}.
\end{split}
\end{equation}
Making $n=k-2$, we obtain
\begin{equation}
\begin{split}
& \left\vert \psi \left( t\right) \right\rangle _{m=1}= \\
& \sum_{n=0}^{\infty }\left\{ i^{n-1}\left[ \frac{1}{\eta t}nJ_{n}\left(
2\eta t\right) -J_{n+1}\left( 2\eta t\right) \right] +i^{n+1}J_{n+3}\left(
2\eta t\right) \right\} .
\end{split}
\end{equation}
Using again the recurrence relation (\ref{6.28}), we finally write \newline
\begin{equation}
\left\vert \psi \left( t\right) \right\rangle _{m=1}=\sum_{n=0}^{\infty }
\left[ i^{n-1}J_{n-1}\left( 2\eta t\right) +i^{n+1}J_{n+3}\left( 2\eta
t\right) \right] \left\vert n\right\rangle .  \label{6.35}
\end{equation}%
Following the same procedure, it is easy to prove that for $m=2$,
\begin{equation}
\left\vert \psi \left( t\right) \right\rangle _{m=2}=\sum_{n=0}^{\infty }
\left[ i^{n-2}J_{n-2}\left( 2\eta t\right) +i^{n+2}J_{n+4}\left( 2\eta
t\right) \right] \left\vert n\right\rangle .  \label{6.36}
\end{equation}
Rewriting results (\ref{6.29}), (\ref{6.35}) and (\ref{6.36}),
\begin{equation}
\begin{split}
& \left\vert \psi \left( t\right) \right\rangle _{m=0}= \\
& =\sum_{n=0}^{\infty }\left[ i^{n-0}J_{n-0}\left( 2\eta t\right)
+i^{n+0}J_{n+0+2}\left( 2\eta t\right) \right] \left\vert n\right\rangle , \\
& \left\vert \psi \left( t\right) \right\rangle _{m=1}= \\
& =\sum_{n=0}^{\infty }\left[ i^{n-1}J_{n-1}\left( 2\eta t\right)
+i^{n+1}J_{n+1+2}\left( 2\eta t\right) \right] \left\vert n\right\rangle , \\
& \left\vert \psi \left( t\right) \right\rangle _{m=2}= \\
& =\sum_{n=0}^{\infty }\left[ i^{n-2}J_{n-2}\left( 2\eta t\right)
+i^{n+2}J_{n+2+2}\left( 2\eta t\right) \right] \left\vert n\right\rangle , \\
& ... \\
& ...
\end{split}
\end{equation}
It is easy to see that the solution for the $m$-initial condition is \newline
\begin{equation}
\begin{split}
& \left\vert \psi \left( t\right) \right\rangle _{m}= \\
& =\sum_{n=0}^{\infty }\left[ i^{n-m}J_{n-m}\left( 2\eta t\right)
+i^{n+m}J_{n+m+2}\left( 2\eta t\right) \right] \left\vert n\right\rangle . \label{6.37}
\end{split}
\end{equation}

We have constructed a new expression for nonlinear coherent states and that
we call \textbf{SG} coherent states. We have managed to construct an
expression that allows us to study the time evolution of \textbf{SG}
coherent states for an arbitrary $\left\vert m\right\rangle $ initial
condition. We also found the physical interpretation of the parameter $x$
(used in previous sections) as a normalized interaction time with respect to
the coupling strength, i.e., $x=\eta t$. We analyze now the nonclassical
features of the constructed states, so we have to make use of the methods
previously mentioned, these are the $\mathcal{Q}$ function, the photon
number distribution and the Mandel $Q$-parameter.

\subsection{Time-dependent \textbf{SG} coherent states analysis}
To perform a complete description of the constructed states (\ref{6.37}), we
have to verify if they present the nonclassical features that nonlinear
coherent states may exhibit. In order to study these nonclassical features,
we propose to use three methods. First, we analyze their behavior in phase
space via the $\mathcal{Q}$ function; then, because we want to analyze
amplitude squeezing and quantum interferences, we show the photon number
distribution of the constructed states, and finally, as we want to know when
the constructed states are maximally squeezed, we show the Mandel $Q$%
-parameter.

\subsubsection{$\mathcal{Q}$ function}
Considering the definition $\mathcal{Q}=\frac{1}{\pi }\left\langle \alpha |
\hat{\rho}|\alpha \right\rangle $ and writing $\hat{\rho}$ in terms of the
constructed states (\ref{6.37}), we write the $\mathcal{Q}$ function for the
time-dependent \textbf{SG} coherent states as
\begin{equation}
\begin{split}
&\mathcal{Q}_\text{SG}\left( \alpha \right) = \\
&\frac{e^{-\left\vert \alpha \right\vert }}{\pi }\left\vert
\sum_{n=0}^{\infty }\frac{\alpha ^{\ast ^{n}}}{\sqrt{ n!}}\left[
i^{n-m}J_{n-m}\left( 2\eta t\right) +i^{n+m}J_{n+m+2}\left( 2\eta t\right)
\right] \right\vert ^{2}.
\end{split}
\end{equation}

\begin{figure}[h] \begin{center}
      \epsfig{file=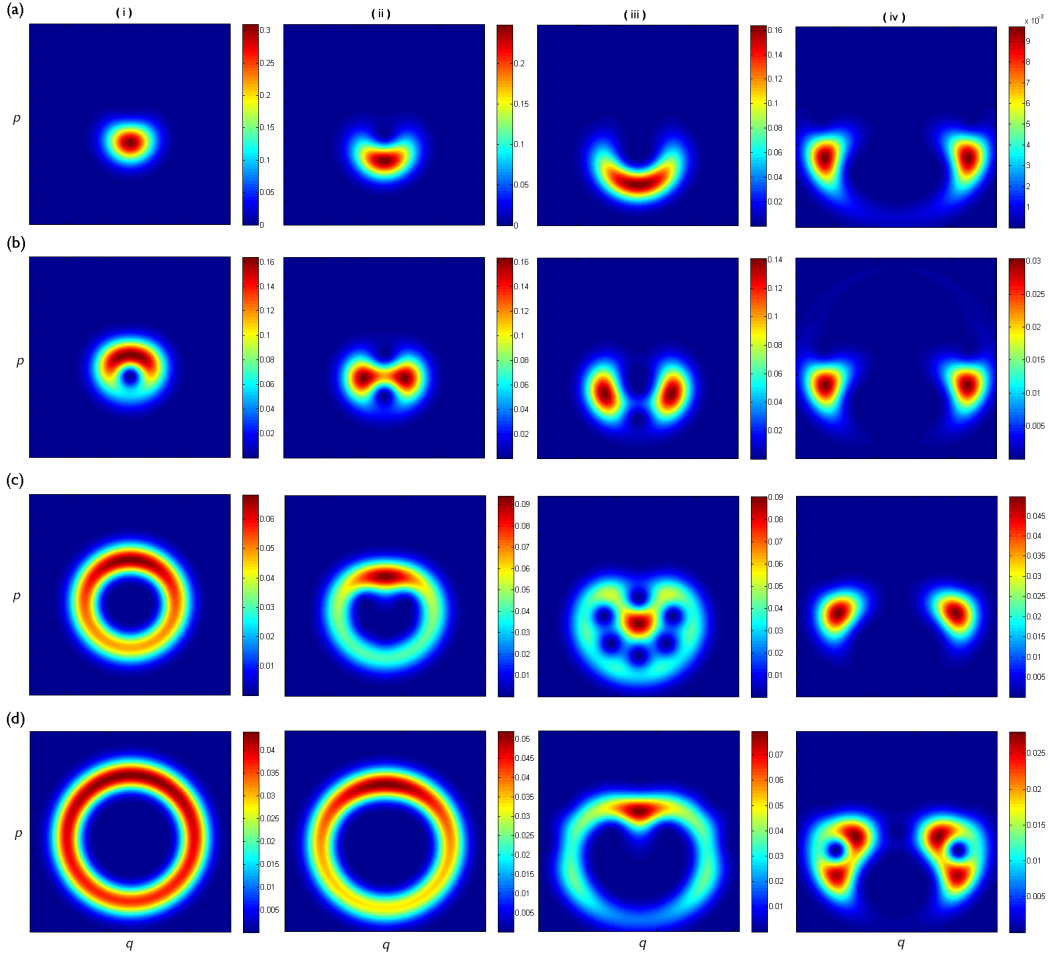,width=14 cm}
\caption{{\protect\footnotesize {\textbf{SG} coherent states }}$\mathcal{Q}$%
{\protect\footnotesize {\ function for (i) $\protect\eta t=1$; (ii) $\protect%
\eta t=2.32$; (iii) $\protect\eta t=5$ and (iv) $\protect\eta t=20$, with
initial conditions (a) $\left\vert 0\right\rangle $;\hspace{1mm} (b) $%
\left\vert 1\right\rangle $;\hspace{1mm} (c) $\left\vert 5\right\rangle $
and (d) $\left\vert 10\right\rangle $.}}}       \label{fig61}
\end{center} \end{figure}

Figure \ref{fig61} shows the time-evolved $\mathcal{Q}$ function of \textbf{SG}
coherent states for different initial conditions. The figure is structured
as follows: The time evolution is shown in each row, for example, (a)
represents the time evolution of \textbf{SG} coherent states for $\left\vert
0\right\rangle $ as initial condition and the index (i) represents the
normalized interaction time $\eta t$. We have chosen (ii) to be $\eta t=2.32
$, because, as we showed previously, it is the time when the state with
initial condition $\left\vert 0\right\rangle $ is maximally squeezed.
The nonclassical features of the constructed states are summarized in Figure
\ref{fig61}. \textbf{SG} coherent states present a strong amplitude squeezing (Figure
\ref{fig61}(a,ii)), splitting into two coherent-like states (Figure \ref{fig61}(b,iii)) and,
as a consequence of the splitting, pronounced quantum interferences (Figure
\ref{fig61}(c,iii)). \newline

An interesting result is that no matter what initial condition we choose,
\textbf{SG} coherent states eventually split into two coherent-like states.
This is very interesting result because such a distribution corresponds to
states called Schr\"{o}dinger's cat states, which are very useful in quantum
information processing \cite{25}.

\subsubsection{Photon number distribution}
\begin{figure}[h] \begin{center}
      \epsfig{file=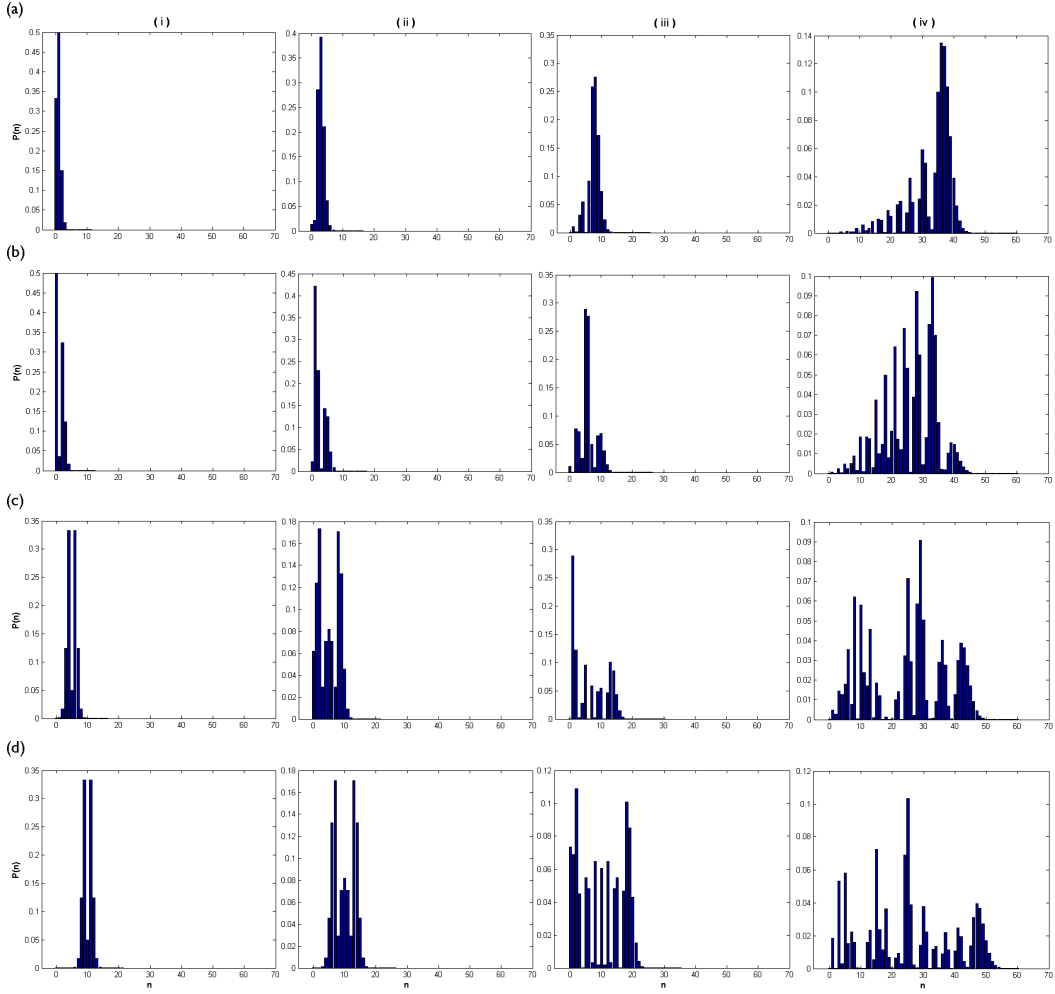,width=14cm}
\caption{{\protect\footnotesize {\textbf{SG} coherent states photon number
probability distributions for (i) $\protect\eta t=1$; (ii) $\protect\eta %
t=2.32$; (iii) $\protect\eta t=5$ and (iv) $\protect\eta t=20$, with initial
conditions (a) $\left\vert 0\right\rangle $;\hspace{1mm} (b) $\left\vert
1\right\rangle $;\hspace{1mm} (c) $\left\vert 5\right\rangle $ \hspace{1mm}
and (d) $\left\vert 10\right\rangle $.}}}
      \label{fig62}
\end{center}  \end{figure}

To complement the description of the nonclassical features that we observed
from the $\mathcal{Q}$ function, we show in Figure \ref{fig62} the photon number distribution of
the \textbf{SG} coherent states considering the same conditions of Figure \ref{fig61}.
\newline
Time-dependent \textbf{SG} coherent states photon number distribution for
the $m$-initial condition is given by
\begin{equation}
P_{m}\left( n,t\right) =\left\vert i^{n-m}J_{n-m}\left( 2\eta t\right)
+i^{n+m}J_{n+m+2}\left( 2\eta t\right) \right\vert ^{2},  \label{6.39}
\end{equation}

\subsubsection{Mandel $Q$-parameter}
\begin{figure}[h] \begin{center}
      \epsfig{file=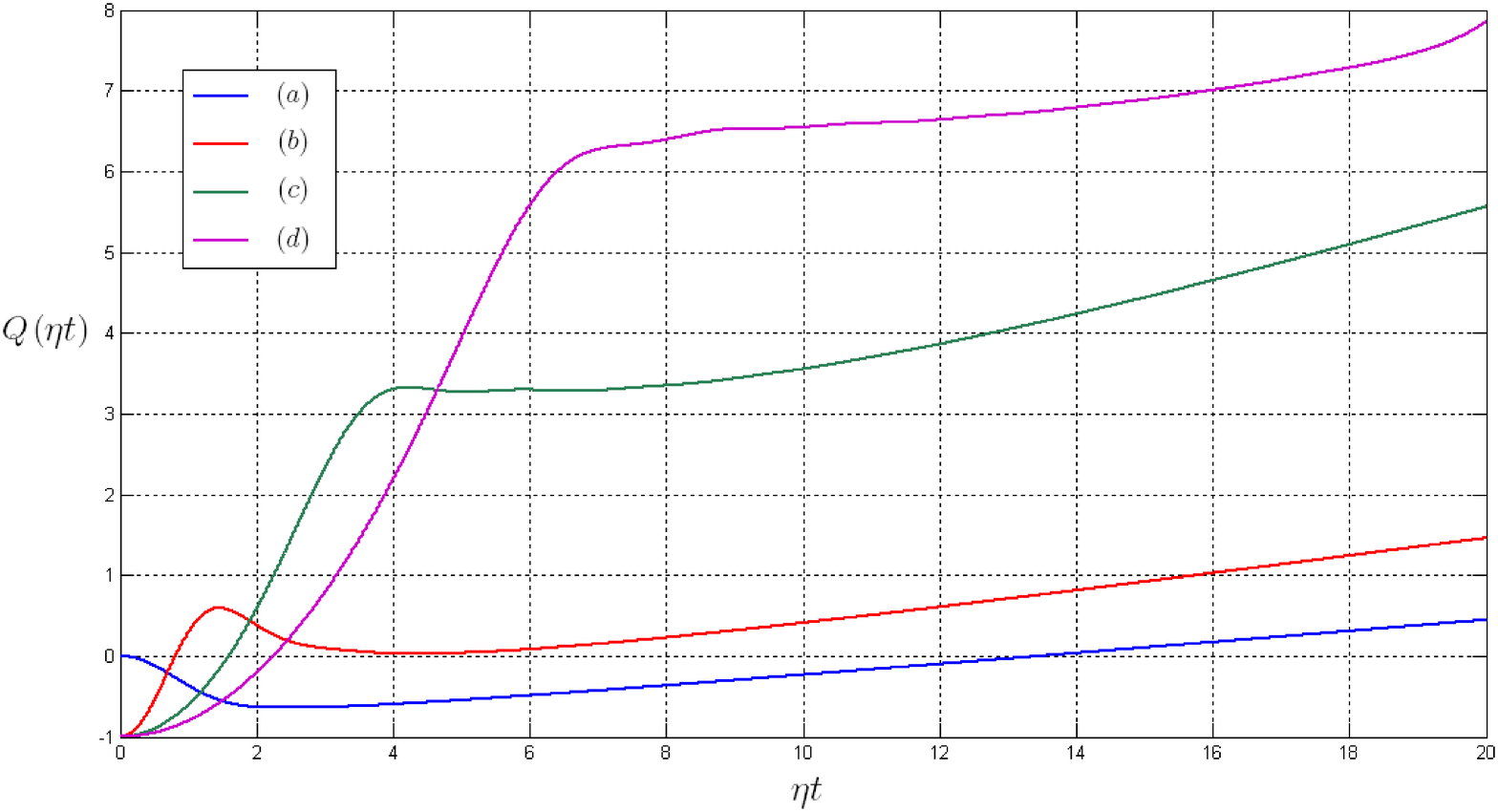,width=12cm}
\caption{{\protect\footnotesize {\textbf{SG} states Mandel Q-parameter with
initial conditions (a) $\left\vert 0\right\rangle $; (b) $%
\left\vert 1\right\rangle $; (c) $\left\vert 5\right\rangle $ and (d) $%
\left\vert 10\right\rangle $. }}}
      \label{fig63}
\end{center}   \end{figure}

As we want to know the time domain for which the constructed states exhibit
amplitude squeezing, and moreover, we want to know when the states are
maximally squeezed, we obtain the Mandel $Q$-parameter for the
time-dependent \textbf{SG} coherent states. \newline
We have that
\begin{equation}
\left\langle \hat{n}\right\rangle =\sum_{n=0}^{\infty }n\left[ J_{n-m}\left(
2\eta t\right) +\left( -1\right) ^{m}J_{n+m+2}\left( 2\eta t\right) \right]
^{2},
\end{equation}
and
\begin{equation}
\left\langle \hat{n}^{2}\right\rangle =\sum_{n=0}^{\infty }n^{2}\left[
J_{n-m}\left( 2\eta t\right) +\left( -1\right) ^{m}J_{n+m+2}\left( 2\eta
t\right) \right] ^{2}.
\end{equation}
\newline
Substituting in the definition of the Mandel $Q$-parameter, equation (\ref%
{5.26}), we obtain the plot shown in Figure \ref{fig63}.
From Figure \ref{fig63} we see that (a) shows the $Q$-parameter for the SG coherent
states where the maximum squeezing happens when $\eta t=2.32$. From the
others, we cannot observe squeezing, because the initial condition is a full
squeezed state, a number state; however, we obtain that an initial number
state eventually transforms into two coherent-like states that we may
identify as Schr\"{o}dinger's cat states and, due to the splitting, we
observe quantum interferences. \newline
The generalization (\ref{6.37}) does not give us different effects from the
ones that we may obtain from (\ref{6.20}); however, as we will see in the
next section, equation (\ref{6.37}) helps us to show that nonlinear coherent
states may be modeled by propagating light in semi-infinite arrays of
optical fibers.

\section{Classical quantum analogies}
The modeling of quantum mechanical systems with classical optics is a topic
that has attracted interest recently. Along these lines Man'ko\textit{\ et
al.}\cite{26} have proposed to realize quantum computation by quantum like
systems, Ch\'avez-Cerda \textit{et al} \cite{27}  have shown how quantum-like entanglement may be realized in classical optics, and Crasser \textit{et al.}\cite{28} have pointed out the
similarities between quantum mechanics and Fresnel optics in phase space.
Following these cross-applications, here we show that nonlinear coherent
states may be modeled by propagating light in semi-infinite arrays of
optical fibers. \newline
Makris \textit{et al.}\cite{29} have shown that for a semi-infinite array of
optical fibers, the normalized modal amplitude in the $n$th optical fiber
(after the $m$th has been initially excited) is written as
\begin{equation}
a_{n}\left( Z\right) =A_{0}\left[ i^{n-m}J_{n-m}\left( 2Z\right)
+i^{n+m}J_{n+m+2}\left( 2Z\right) \right] ,  \label{6.42}
\end{equation}
where $Z=cz$ is the normalized propagation distance with respect to the
coupling coefficient $c$. \newline
We see that, for $A_{0}=1$ the normalized intensity distribution is
\begin{equation}
I_{n}\left( Z\right) =\left\vert i^{n-m}J_{n-m}\left( 2Z\right)
+i^{n+m}J_{n+m+2}\left( 2Z\right) \right\vert ^{2}.  \label{6.43}
\end{equation}
Rewriting (\ref{6.39}) and using the normalized interaction time with
respect to the coupling coefficient $\eta $, i.e., $x=\eta t$, we have
\begin{equation}
P_{m}\left( n,x\right) =\left\vert i^{n-m}J_{n-m}\left( 2x\right)
+i^{n+m}J_{n+m+2}\left( 2x\right) \right\vert ^{2}.  \label{6.44}
\end{equation}
As equations (\ref{6.43}) and (\ref{6.44}) are the same, we conclude that
the photon number distribution for the \textbf{SG} coherent states may be
modeled by the intensity distribution of propagating light in semi-infinite
arrays of optical fibers. We have found a new relation between quantum
mechanical systems and classical optics.

\section{Conclusions}
We managed to construct a new expression for nonlinear coherent states, that
we called \textbf{SG} coherent states, by developing the displacement
operator in a Taylor series. The importance of nonlinear coherent states
resides in the nonclassical features that they may exhibit and, in order to
analyze the nonclassical behavior of the constructed states, we presented
three methods: the $\mathcal{Q}$ function, the photon number distribution
and the Mandel $Q$-parameter. \newline

The expression for the \textbf{SG} coherent states depends on a parameter $x$
and, considering the values of $x$, we found that
\begin{itemize}
\item For a certain domain of $x$, \textbf{SG} coherent states exhibit
amplitude squeezing; i.e., their photon number distribution is narrower than
the one of a coherent state of the same amplitude. Using the Mandel $Q$%
-parameter we determined that maximum squeezing occurs when $x=2.32$.
\item For larger values of $x$, \textbf{SG} coherent states split into two
coherent-like states, this superposition of coherent states is called Schr%
\"{o}dinger's cat states and are very useful in quantum information
processing \cite{25}. We also found that the splitting of the states gives
rise to quantum interference effects.\newline
\end{itemize}

We also constructed \textbf{SG} coherent states as eigenfunctions of a
Hamiltonian for the \textbf{SG} operators representing physical couplings to
the radiation field. We managed to obtain the solution as the one obtained
before, except for a phase term that rotates by $\pi /2$ the $\mathcal{Q}$
function, but it does not change the results that we have already found. The
obtained expression for \textbf{SG} coherent states helped us to physically
interpret the parameter $x$ as a normalized interaction time with respect to
the coupling coefficient $\eta $. \newline

The formalism presented in this part allowed us to construct a solution for
an arbitrary initial condition $\left\vert m\right\rangle $. Equation (\ref%
{6.37}) is the most important result of this work, because it represents a
new expression for nonlinear coherent states; moreover, it represents a
general expression for an arbitrary $\left\vert m\right\rangle $ initial
condition. \newline

Analyzing expression (\ref{6.37}), we found that a multiple splitting of the
states occurs; this gives rise to interesting structures of the phase-space
distribution as the one shown in Figure \ref{fig61}(c,iii). Except for that result,
the generalized solution presented the same effects as the particular case $%
m=0$; however, it turns out to be very useful to establish an interesting
relation between quantum and classical optics. We showed that \textbf{SG}
coherent states photon number distribution may be modeled by the intensity
distribution of propagating light in semi-infinite arrays of optical fibers.
With this finding, we have presented a new analogy between quantum
mechanical systems and classical optics.

\section*{Appendix}
\appendix
\section*{Sums of the Bessel functions of the first kind of integer order}
We will derive in this appendix the solution of some sums of Bessel functions of the first kind of integer order that appear in several applications, and in particular, that appear in this contribution in the section where we calculate the Mandel $Q-$parameter (section 4.3).
We will demonstrate that
\begin{equation}\label{b2.10} \begin{split}
    &\sum\limits_{k=1}^{\infty }k^{2\nu }J_{k}^{2}\left( x\right)= \\
    & =\dfrac{\left(-1\right) ^{\nu }}{4\pi }\int\limits_{-\pi }^{\pi }B_{2\nu }\left(
g^{\prime }\left( y\right) ,g^{\prime \prime }\left( y\right) ,...,g^{\left(
2\nu \right) }\left( y\right) \right) dy,
\end{split}\end{equation}
where $\nu$ is a positive integer, $g\left( y\right) =ix\sin y$ \ and $B_{n}\left(
x_{1},x_{2},...,x_{n}\right) $ is the complete Bell polynomial \cite{30,31,32} given by the
following determinant:
\begin{spacing}{1.8}
\begin{equation}  \begin{split}\label{}
    &B_n(x_1,x_2,...,x_n)= \\
    &=\det \left|
    \begin{array}{cccccccc}
      x_1&\binom{n-1}{1}x_2 &\binom{n-1}{2}x_3 &\binom{n-1}{3}x_4 &\binom{n-1}{4}x_5 & \cdots & \cdots & x_n \\
      -1&x_1 &\binom{n-2}{1}x_2 &\binom{n-2}{2}x_3 &\binom{n-2}{3}x_4 &\cdots &\cdots & x_{n-1} \\
      0&-1 &x_1 &\binom{n-3}{1}x_2 &\binom{n-3}{2}x_3 &\cdots &\cdots &x_{n-2} \\
      0& 0 &-1 & x_1 &\binom{n-4}{1}x_2 & \cdots & \cdots & x_{n-3} \\
      0& 0 &0 & -1 & x_1 & \cdots & \cdots & x_{n-4} \\
      0& 0 & 0 & 0 & -1 & \cdots & \cdots & x_{n-5} \\
      \vdots & \vdots & \vdots & \vdots & \vdots & \ddots & \ddots & \vdots \\
      0& 0 & 0 & 0 & 0 & \cdots & -1 & x_1
    \end{array} \right|.
\end{split} \end{equation}
\end{spacing}
To demonstrate (\ref{b2.10}), we will need the well known Jacobi-Anger expansions for the Bessel functions of the first kind
(\cite{33}, page 933, \cite{34}, page 361; \cite{19}, page 70),
\begin{equation}\label{ja1}
    e^{ix\cos y}=\sum_{n=-\infty}^{\infty}i^nJ_n(x)e^{iny}
\end{equation}
and
\begin{equation}\label{ja2}
    e^{ix\sin y}=\sum_{n=-\infty}^{\infty}J_n(x)e^{iny}.
\end{equation}
Using expression (\ref{ja1}), we can easily write
\begin{equation}
    \dfrac{d^{n}}{dy^{n}}e^{ix\sin y}=i^{n}\sum\limits_{k=-\infty }^{\infty }k^{n}J_{k}\left(x\right) e^{iky}.
\end{equation}
To calculate the $n$-derivative in the left side of equation above, we use the Fa\`{a} di Bruno's formula (\cite{33}, page 22) for the $n$-derivative of the composition
\begin{equation}\label{faa}  \begin{split}
    & \dfrac{d^{n}}{dx^{n}}f\left( g\left( x\right) \right)= \\
    &=\sum\limits_{k=0}^{n}f^{\left( k\right) }\left( g\left( x\right) \right)
\cdot B_{n,k}\left( g^{\prime }\left( x\right) ,g^{\prime \prime }\left(
x\right) ,...,g^{\left( n-k+1\right) }\left( x\right) \right),
\end{split} \end{equation}
where $B_{n,k}\left( x_{1},x_{2},...,x_{n-k+1}\right) $ is a Bell polynomial \cite{30,31,32}, given by
\begin{equation}  \begin{split}
    &B_{n,k}\left( x_{1},x_{2},...,x_{n-k+1}\right)= \\
    &=\sum \dfrac{n!}{j_{1}!j_{2}!...j_{n-k+1}!}   \\
    &\times \left( \dfrac{x_{1}}{1!}\right) ^{j_{1}}\left(
\dfrac{x_{2}}{2!}\right) ^{j_{2}}...\left( \dfrac{x_{n-k+1}}{\left(
n-k+1\right) !}\right) ^{j_{n-k+1}},
\end{split} \end{equation}
the sum extending over all sequences $j_{1},j_{2},j_{3},...,j_{n-k+1}$ of non-negative integers such that
$j_{1}+j_{2}+...+j_{n-k+1}=k$ and $j_{1}+2j_{2}+3j_{3}+...+\left(n-k+1\right) j_{n-k+1}=n.$

Using (\ref{faa}),
\begin{equation}\begin{split}
    &\dfrac{d^{n}}{dy^{n}}e^{ix\sin y}= \\
    &=e^{ix\sin y}\sum\limits_{k=0}^{n}B_{n,k}\left( g^{\prime }\left(
    x\right) ,g^{\prime \prime }\left( x\right) ,...,g^{\left( n-k+1\right)}\left( x\right) \right).
\end{split} \end{equation}
We multiply now for the complex conjugate of (\ref{ja2}) and obtain
\begin{equation}\begin{split}
    &i^{n}\sum\limits_{k,l=-\infty }^{\infty }k^{n}J_{l}\left( x\right)
J_{k}\left( x\right) e^{i\left( k-l\right) y}= \\
&=B_{n}\left( g^{\prime }\left(
y\right) ,g^{\prime \prime }\left( y\right) ,...,g^{\left( n\right) }\left(
y\right) \right).
\end{split} \end{equation}
Integrating both sides of the above equation from $-\pi$ to $\pi$, and using that $\int_{-\pi}^{\pi}e^{i(k-l)y}dy=\delta_{kl}$, we arrive to the formula we wanted
\begin{equation}\ \begin{split}
    &\sum\limits_{k=1}^{\infty }k^{2\nu }J_{k}^{2}\left( x\right) =  \\
    &=\dfrac{\left(
-1\right) ^{\nu }}{4\pi }\int\limits_{-\pi }^{\pi }B_{2\nu }\left(
g^{\prime }\left( y\right) ,g^{\prime \prime }\left( y\right) ,...,g^{\left(
2\nu \right) }\left( y\right) \right) dy.
\end{split} \end{equation}

In particular, as the complete Bell polynomials for $n=2$ and $n=4$, are
\begin{equation}\label{}
    B_2(x_1,x_2)=x_1^2+x_2
\end{equation}
and
\begin{equation}\label{}
    B_4(x_1,x_2,x_3,x_4)=x_1^4+6x_1^2x_2+4x_1x_3+3x_2^2+x_4,
\end{equation}
it is very easy to show that,
\begin{equation}\label{}
  \sum\limits_{k=1}^{\infty}k^{2}J_{k}^{2}\left( x\right) =\dfrac{1}{4}x^{2}
\end{equation}
and
\begin{equation}\label{}
\sum\limits_{k=1}^{\infty}k^{4}J_{k}^{2}\left( x\right) =\allowbreak \dfrac{3}{16}x^{4}+\dfrac{1}{4}x^{2}.
\end{equation}

\end{document}